\documentclass{article}
\usepackage[margin=1in]{geometry}
\usepackage{amsmath}
\usepackage{epsfig}

\numberwithin{equation}{section} 

\newcommand{\beq}{\begin{equation}}
\newcommand{\eeq}{\end{equation}}
\newcommand{\beqa}{\begin{eqnarray}}
\newcommand{\eeqa}{\end{eqnarray}}
\newcommand{\lslash}[1]{#1\llap/}

\newcommand{\Eq}[1]{Eq.\ (\ref{#1})}
\newcommand{\Eqs}[2]{Eqs.\ (\ref{#1}) and (\ref{#2})}
\newcommand{\Eqss}[3]{Eqs.\ (\ref{#1}), (\ref{#2}) and (\ref{#3})}
\newcommand{\Ref}[1]{Ref.\ \cite{#1}}
\newcommand{\Refs}[2]{Refs.\ \cite{#1} and \cite{#2}}

\newcommand{\Fig}[1]{Fig.\ \ref{#1}}
\newcommand{\Section}[1]{Section\ \ref{#1}}
\newcommand{\Sections}[2]{Section\ \ref{#1} and \ref{#2}}
\newcommand{\Appendix}[1]{Appendix\ \ref{#1}}

\title{
  Neutrino damping in a fermion and scalar background
}
  
\author{Jos\'e F. Nieves\footnote{nieves@ltp.uprrp.edu}\\
  Laboratory of Theoretical Physics, Department of Physics\\
  University of Puerto Rico, R\'{\i}o Piedras, Puerto Rico 00936
  \and\\[12pt]
  Sarira Sahu\footnote{sarira@nucleares.unam.mx}\\
  Instituto de Ciencias Nucleares\\
  Universidad Nacional Aut\'onoma de Mexico\\
  Circuito Exterior, C. U.\\
  A. Postal 70-543, 04510 Mexico DF, Mexico\\
}
\date{}

\begin{document}
\maketitle

\begin{abstract}
  We consider the propagation of a neutrino
  in a background composed of a scalar particle and a fermion
  using a simple model for the coupling of the form $\lambda\bar f_R\nu_L\phi$.
  In the presence of these interactions there can be damping terms
  in the neutrino effective potential and index of refraction.
  We calculate the imaginary part of the neutrino self-energy in this case,
  from which the damping terms are determined.
  The results are useful in the context of Dark Matter-neutrino interaction
  models in which the scalar and/or fermion constitute the dark-matter.
  The corresponding formulas for models in which the scalar particle
  couples to two neutrinos via a coupling of the form
  $\lambda^{(\nu\nu\phi)}\bar\nu^c_R\nu_L\phi$
  are then obtained as a special case, which
  can be important also in the context of neutrino collective
  oscillations in a supernova and in the Early Universe hot plasma
  before neutrino decoupling.
  A particular feature of our results is that the damping term
  in a $\nu\phi$ background is independent of the
  antineutrino-neutrino asymmetry in the background.
  Therefore, the relative importance of the damping term
  may be more significant if the neutrino-antineutrino asymmetry
  in the background is small, because the leading 
  $Z$-exchange and $\phi$-exchange contributions to the effective potential,
  which are proportional to the neutrino-antineutrino asymmetry,
  are suppressed in that case, while the damping term is not.
\end{abstract}

\section{Introduction and Summary}

Many extensions of the standard electroweak theory that have been
considered recently involve neutrino interactions with scalar particles
of the form
\beq
\label{Lnunuphi}
L_{int}= \frac{\lambda^{(\nu\nu\phi)}}{2}{\bar \nu}^c_R {\nu}_L\phi + h.c.
\eeq
Constraints on the properties and interactions of such scalar particles
as well as their possible effects have been studied in the context
of particle physics, astrophysics, and cosmology\cite{Berryman:2018ogk,
Farzan:2018gtr,
Stephenson:1993rx,hm:2017mio,Heurtier:2016otg, 
Sawyer:2006ju,Pasquini:2015fjv,Ge:2017poy,Brdar:2017kbt,Ohlsson:2012kf,
Antusch:2008tz, Aartsen:2017xtt}.

In a previous work\cite{Nieves:2018vxl} we noted
that such couplings can produce
nonstandard contributions to the neutrino index
of refraction and effective potential when the neutrino propagates
in a neutrino background. This occurs
in the environment of a supernova, where it is now well known that
the effect leads to the collective neutrino oscillations
and related phenomena (see for example \Refs{Duan:2010bg}{Chakraborty:2016yeg}
and the works cited therein), and it can also occur in the hot
plasma of the Early Universe before the neutrinos
decouple\cite{Wong:2002fa,Mangano:2006ar}. The presence of such scalars
and the $\nu\nu\phi$ couplings can have effects in those contexts.

In \Ref{Nieves:2018vxl} we considered the real part of the self-energy
of a neutrino that propagates in a medium consisting of fermions and scalars,
with a coupling of the form
\beq
\label{Lfnuphi}
L_{int} = \lambda\bar f_R \nu_L \phi + h.c.\,.
\eeq
Such couplings have been considered recently in the context of
Dark Matter-neutrino interactions\cite{Primulando:2017kxf,Franarin:2018gfk,
Mangano:2006mp,Binder:2016pnr,Campo:2017nwh,Brune:2018sab}.
From the self-energy, the neutrino and antineutrino effective potential and
dispersions relation were then determined.
The corresponding formulas for the case of the neutrino and scalar background,
with the couplings given in \Eq{Lnunuphi}, are obtained as the special
case in which $f_R \rightarrow \nu^c_R$.

Our motivation in the present work is the fact that in the presence of these
interactions there can be damping terms in the neutrino effective
potential and index of refraction. Such damping terms
arise as a consequence of processes such as
$\nu + \phi \leftrightarrow f$ and $\nu + \bar f\leftrightarrow \bar\phi$,
that become possible and affect the neutrino propagation
depending on the kinematics conditions.
Here we extend our previous work to calculate the imaginary
part of the neutrino self-energy, from which
the damping terms in the neutrino effective potential
and dispersion relations are determined.
To be specific, we calculate the imaginary
part (or more precisely the absorptive part) of the neutrino self-energy, 
in a fermion and scalar background due to the interaction 
given in \Eq{Lfnuphi}. From the imaginary part of the 
self-energy the damping terms in the effective potential and dispersion
relation are then obtained. 
The formulas for the case of the neutrino and scalar background,
with the couplings given in \Eq{Lnunuphi} are obtained as the special
case in which $f_R \rightarrow \nu^c_R$. We obtain explicit formulas
for the damping terms in the case that the background particle momentum
distributions have the standard isotropic form. But for generality
and in order to present the results in a way that can be useful
to the situations described and others not considered here, we
consider the case of an anisotropic background as well.

As in our previous work, in writing \Eq{Lnunuphi} we have taken
into account only the diagonal neutrino coupling and assume
the presence of only one scalar field.
In the more general case, with more neutrino
species in the background and non-diagonal neutrino-$\phi$ couplings,
the density matrix formalism\cite{Duan:2010bg,Wong:2002fa}
would have to be used which is beyond the scope of our work.
Nevertheless, despite the simplification made by considering
only one neutrino type, our results illustrate some features that can
serve as a guide when considering more general cases
or situations not envisioned here. For example, in the context
of models in which sterile neutrinos have \emph{secret} gauge
interactions of the form
$\bar\nu_s\gamma^\mu \nu_s A^\prime_\mu$\cite{nusterilesecret},
similar considerations apply 
when a sterile neutrino propagates in a background of sterile
neutrinos and $A^\prime$ bosons. Similar effects arise in models in which
the sterile neutrino  and an active neutrino have an interaction of the form
$\lambda^{(\nu\nu_s\phi)}{\bar \nu}^c_{Rs} {\nu}_L\phi$, when
either $\nu$ or $\nu_s$ propagates in a background of the other.
The formulas we obtain for the damping terms
can be applied in the context of such models with minor modifications.

A particular feature of our results is that the damping term
in a $\nu\phi$ background can be as large
as the $\nu\phi$ contribution to the neutrino effective potential. Moreover,
in contrast to the latter, it is independent of the
antineutrino-neutrino asymmetry $(n_{\nu} - n_{\bar\nu})$
in the background. Therefore, the relative importance of the damping term
may be more significant if the neutrino-antineutrino asymmetry
in the background is small, because the leading 
$Z$-exchange and $\phi$-exchange contributions to the effective potential
are suppressed in that case, while the damping term is not.

In \Section{sec:selfenergy} we explain the procedure we follow
to calculate the imaginary part of the neutrino thermal self-energy,
from which the damping term in the dispersion relation and effective
potential is determined. The calculation of the damping term
in a background of $f$ and $\phi$ particles
is carried out in \Section{sec:dampingtermcalculation}.
The imaginary part of the neutrino effective potential is expressed
as an integral over the momentum distribution functions
of the background particles and each term can be identified
with the corresponding process that contributes
to the neutrino damping. The formula for the
damping term in the neutrino and antineutrino
dispersion relations is then obtained. It is expressed as a set of
integrals over the distribution functions, which depend on the 
kinematic conditions in the situation being considered. The integral
formulas are evaluated explicitly for the standard thermal distributions
but, as we explain, they could be used in more general cases as well.
In order to illustrate some of the main features of the results obtained,
in \Section{sec:discussion} we consider various example situations,
such as a degenerate $f$ background or a classical relativistic $\phi$
background, and evaluate the damping term explicitly in each case.
There we also consider the case of a neutrino propagating in a
$\nu\phi$ background. We briefly summarize our results
in \Section{sec:conclusions}.
Some of the details regarding the evaluation of the integrals
in \Section{sec:dampingtermcalculation} are provided in the Appendix.

\section{Imaginary part of the self-energy and dispersion relation}
\label{sec:selfenergy}

\subsection{Self-energy}
\label{subsec:selfenergy}

We consider the interaction given in \Eq{Lfnuphi}
%
%
and calculate the neutrino damping in a background composed of
$f$ and $\phi$ particles. For definiteness we assume that the neutrino
is massless, but we consider $m_{f,\phi}$ to be non-zero in general.
The results can be applied to the situation in which a
neutrino propagates through a $\nu\phi$
background, with the interaction as given in \Eq{Lnunuphi},
as a special case by putting $f_R\rightarrow \nu^c_R$
in the resulting formulas.

For the calculation of the damping term in the dispersion relation
we follow the work of \Ref{DOlivo:1993rhs}. We summarize below
the necessary aspects needed, adapted to the present situation.
We denote by $u^\mu$ the velocity four-vector of the background medium
and by $k^\mu$ the momentum of the propagating neutrino. In the
background medium's own rest frame,
\beq
u^\mu = (1,\vec 0)\,,
\eeq
and in that frame we write the components of $k^\mu$ in the form
\beq
\label{komegakappa}
k^\mu = (\omega,\vec\kappa)\,.
\eeq
The dispersion relation $\omega(\vec\kappa)$ and spinor of the propagating
mode are given by the solutions of
\beq
\label{fieldeq}
\left(\lslash{k} - \Sigma_{eff}\right)\psi_L(k) = 0\,,
\eeq
where $\Sigma_{eff}$ is the effective thermal self-energy of the
propagating neutrino.  It can be decomposed into its dispersive
and absorptive parts $\Sigma_{r,i}$ in the form
\beq
\Sigma_{eff} = \Sigma_r + i\Sigma_i\,,
\eeq
where
\beqa
\Sigma_r & = & \frac{1}{2}\left(\Sigma_{eff} + \bar\Sigma_{eff}\right)\,,
\nonumber\\
\Sigma_i & = & \frac{1}{2i}\left(\Sigma_{eff} - \bar\Sigma_{eff}\right)\,,
\eeqa
with
\beq
\bar\Sigma_{eff} = \gamma^0\Sigma^\dagger_{eff}\gamma^0 \,.
\eeq
The dispersive part $\Sigma_r$ is obtainable from the formula,
\beq
\Sigma_r = \Sigma_{11r} \equiv \frac{1}{2}
\left(\Sigma_{11} + \bar\Sigma_{11}\right)\,,
\eeq
where $\Sigma_{11}(k)$ is 11 component of the neutrino thermal
self-energy matrix. The calculation of $\Sigma_r$ was the
subject of \Ref{Nieves:2018vxl}.

\begin{figure}
\begin{center}
\epsfig{file=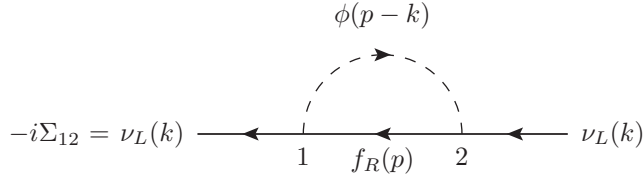,bbllx=73,bblly=329,bburx=534,bbury=443}
\end{center}
\caption{
  One-loop diagram for the $\Sigma_{12}$ element of the neutrino thermal
  self-energy in an $f\phi$ background.
  \label{fig:sigma12fphi}
}
\end{figure}

Our focus here is on $\Sigma_i$. As emphasized in \Ref{DOlivo:1993rhs},
a convenient way to calculate it is via the formula
\beq
\label{Sigmaidef}
\Sigma_i = \frac{\Sigma_{12}}{2in_F(x)}\,,
\eeq
where $\Sigma_{12}(k)$ is the $12$ element of the neutrino
thermal self-energy matrix, which is calculated by
the diagram displayed in \Fig{fig:sigma12fphi}.
Here we have defined
\beq
x = \beta k\cdot u - \alpha\,,
\eeq
and
\beq
\label{nzfermion}
n_F(z) = \frac{1}{e^z + 1}\,,
\eeq
is the fermion distribution function, written in terms of a dummy variable $z$.
For future reference we define here the corresponding function for the
bosons,
\beq
\label{nzboson}
n_B(z) = \frac{1}{e^z - 1}\,.
\eeq

\subsection{Dispersion relation - damping term}

The question we address here is how to determine the damping term
in the dispersion relation, given the result for $\Sigma_i$.
As already mentioned, the dispersion relation is determined by
solving \Eq{fieldeq}.
%
%
Furthermore, the chirality of the neutrino interactions imply
that $\Sigma_{eff}$ is of the form\footnote{%
This is strictly true in the massless neutrino limit
that we are considering, as we have already stated in
\Section{subsec:selfenergy}. In practice
it is a valid approximation in the limit that the neutrino
mass can be negeclected in the calculation of the diagrams
involved.}
\beq
\label{Sigmachiral}
\Sigma_{eff} = \lslash{V}L\,.
\eeq
We consider now the determination of the dispersion relation for the case
in which $V^\mu$ contains an absorptive part in addition to the dispersive
part that we have considered in \Ref{Nieves:2018vxl}. Therefore, here we assume
that $V^\mu$ has a real and an imaginary part,
\beq
V^\mu = V^\mu_r + iV^\mu_i\,,
\eeq
We should keep in mind that, in general, both $V^\mu_{r,i}$ are
functions of $\omega$ and $\vec\kappa$, which will indicate by writing
$V^\mu_{r,i}(\omega,\vec\kappa)$ when needed.

We summarize the results as follows. Writing the neutrino and
antineutrino dispersion relations in the form
%
%
\beq
\label{disprelform}
\omega^{(\nu,\bar\nu)} = \omega^{(\nu,\bar\nu)}_r -
\frac{i\gamma^{(\nu,\bar\nu)}}{2}\,,
\eeq
$\omega^{(\nu,\bar\nu)}_r$ is given by
\beq
\label{nudisprelreal}
\omega^{(\nu,\bar\nu)}_r = \kappa + V^{(\nu,\bar\nu)}_{eff}
\eeq
where $V^{(\nu,\bar\nu)}_{eff}$ are the effective potentials
\beqa
\label{Veff}
V^{(\nu)}_{eff} & = & n\cdot V_r(\kappa,\vec\kappa) =
 V^0_r(\kappa,\vec\kappa) - \hat\kappa\cdot\vec V_r(\kappa,\vec\kappa)
\,,\nonumber\\
V^{(\bar\nu)}_{eff} & = & -n\cdot V_r(-\kappa,-\vec\kappa) =
 -V^0_r(-\kappa,-\vec\kappa) + \hat\kappa\cdot\vec V_r(-\kappa,-\vec\kappa)\,,
\eeqa
with
\beq
\label{nmu}
n^\mu = (1,\hat\kappa)\,.
\eeq
On the other hand, for the imaginary part we have
%
%
%
\beqa
\label{nudisprelimg}
-\frac{\gamma^{(\nu)}(\vec\kappa)}{2} & = & 
\frac{n\cdot V_i(\kappa,\vec\kappa)}
{1 - n\cdot\left.
\frac{\partial V_r(\omega,\vec\kappa)}{\partial\omega}
\right|_{\omega = \kappa}}\,,\nonumber\\
-\frac{\gamma^{(\bar\nu)}(\vec\kappa)}{2} & = & 
\frac{n\cdot V_i(-\kappa,-\vec\kappa)}
{1 - n\cdot\left.
\frac{\partial V_r(\omega,-\vec\kappa)}{\partial\omega}
\right|_{\omega = -\kappa}}\,,
\eeqa
where $n^\mu$ is defined in \Eq{nmu}.
\Eqs{nudisprelreal}{nudisprelimg}, allow us
to obtain the neutrino and antineutrino dispersion relation and
damping from the self-energy. In those cases in which the
correction due to the $n\cdot\partial V_r(\omega,\vec\kappa)/\partial\omega$
in the denominator can be neglected, the formulas in \Eq{nudisprelimg}
further simplify to
\beqa
\label{nudisprelimg-simple}
-\frac{\gamma^{(\nu)}(\vec\kappa)}{2} & = & 
n\cdot V_i(\kappa,\vec\kappa)\,,\nonumber\\
-\frac{\gamma^{(\bar\nu)}(\vec\kappa)}{2} & = & 
n\cdot V_i(-\kappa,-\vec\kappa)\,.
\eeqa
%
%
%
%
The formulas in \Eq{Veff} were used in \Ref{Nieves:2018vxl} to calculate
the real part of the neutrino dispersion relation in the $f\phi$
background. \Eq{nudisprelimg-simple} are the
ones we will use in the present work to calculate
the damping terms in the dispersion relation.

These results are obtained as follows.
From \Eqs{fieldeq}{Sigmachiral}, the dispersion relations must satisfy
\beq
(k - V)^2 = 0\,,
\eeq
or
\beq
\label{dispreleq}
\omega - V^0 = \pm\sqrt{(\vec\kappa - V)\cdot(\vec\kappa - \vec V)}\,.
\eeq
Remembering that $V^\mu$ is a function of $\omega$ and $\vec\kappa$,
this is an \emph{implicit} equation that determines the dispersion relations
$\omega(\vec\kappa)$. We consider the solution of \Eq{dispreleq} for the case
in which $V^\mu$ is smaller than $\kappa$, so that we can
treat the equation perturbatively.

Let us (re)consider first the real case ($V^\mu_i = 0$). Denoting
the solutions in this case by $\omega^{(\lambda)}_r$ ($\lambda = \pm$),
they must satisfy
\beqa
\label{disperleqfirstorder}
\omega^{(\lambda)}_r - V^0_r(\omega^{(\lambda)}_r,\vec\kappa) & = &\lambda
\sqrt{\left(\vec\kappa - \vec V_r(\omega^{(\lambda)}_r,\vec\kappa)\right)
\cdot\left(\vec\kappa - \vec V_r(\omega^{(\lambda)}_r,\vec\kappa)\right)}
\nonumber\\
& \simeq & \lambda\left(
\kappa - \hat\kappa\cdot\vec V_r(\omega^{(\lambda)}_r,\vec\kappa)
\right)\,.
\eeqa
Thus, to the lowest order,
\beq
\omega^{(\pm)}_r(\vec\kappa) = V^0_r(\pm\kappa,\vec\kappa) 
\pm\left(\kappa - \hat\kappa\cdot\vec V_r(\pm\kappa,\vec\kappa)\right)\,.
\eeq
The neutrino and antineutrino dispersion relations are given by
\beqa
\omega^{(\nu)}_r & = & \omega^{(+)}_r(\vec\kappa)\,,\nonumber\\
\omega^{(\bar\nu)}_r & = & -\omega^{(-)}_r(-\vec\kappa)\,,
\eeqa
which reduce to \Eq{nudisprelreal}.
%
%
%
%
%
%
%
%

We now generalize this to the case with $V^\mu_i \not= 0$. In analogy
to \Eq{disperleqfirstorder}, from \Eq{dispreleq} we obtain to first order
\beq
\label{dispreleqfirstordercomplex}
\omega^{(\lambda)} - V^0(\omega^{(\lambda)},\vec\kappa)
\simeq \lambda\left(
\kappa - \hat\kappa\cdot\vec V(\omega^{(\lambda)},\vec\kappa)
\right)\,.
\eeq
We seek the solution by writing
\beq
\omega^{(\lambda)} = \omega^{(\lambda)}_r + i\omega^{(\lambda)}_i\,,
\eeq 
and we work under the assumption that the imaginary parts are smaller
than the real parts, in other words
$\omega^{(\lambda)}_i \ll \omega^{(\lambda)}_r$. Therefore we write
\beqa
\label{Vexpansion}
V^\mu(\omega^{(\lambda)},\vec\kappa) & \simeq & 
V^\mu(\omega^{(\lambda)}_r,\vec\kappa)
+ i\omega^{(\lambda)}_i\left.\frac{\partial V^\mu(\omega,\vec\kappa)}
{\partial\omega}\right|_{\omega = \omega^{(\lambda)}_r}\nonumber\\
&&\mbox{}\simeq
V^\mu_r(\omega^{(\lambda)}_r,\vec\kappa) +
iV^\mu_i(\omega^{(\lambda)}_r,\vec\kappa)
+ i\omega^{(\lambda)}_i\left.\frac{\partial V^\mu_r(\omega,\vec\kappa)}
{\partial\omega}\right|_{\omega = \omega^{(\lambda)}_r}\,,
\eeqa
neglecting terms of the second order in the imaginary parts.

Substituting \Eq{Vexpansion} into \Eq{dispreleqfirstordercomplex},
the real part yields \Eq{disperleqfirstorder}, while the
imaginary part, after some rearrangements, yields
\beq
\label{disperleqfirstorderimg}
\omega^{(\lambda)}_i(\vec\kappa) =
\frac{n^{(\lambda)}\cdot V_i(\omega^{(\lambda)}_r,\vec\kappa)}
{1 - n^{(\lambda)}\cdot\left.
\frac{\partial V_r(\omega,\vec\kappa)}{\partial\omega}
\right|_{\omega = \omega^{(\lambda)}_r}}\,,
\eeq
where we have defined
\beq
n^{(\lambda)\mu} = (1,\lambda\hat\kappa)\,.
\eeq
\Eq{disperleqfirstorderimg} can be simplified by
approximating $\omega_r^{(\pm)} \simeq \pm\kappa$ in the evaluation of the
right-hand side. Then, remembering that\footnote{%
  Notice that the antineutrino damping term is not defined with
  the additional minus sign as it is the case for the real part of
  the dispersion relation.
}
\beqa
\omega^{(\nu)}_i(\vec\kappa) & = & \omega^{(+)}_i(\vec\kappa)\nonumber\\
\omega^{(\bar\nu)}_i(\vec\kappa) & = & \omega^{(-)}_i(-\vec\kappa)\,,
\eeqa
and identifying the damping rate by
\beq
\omega^{(\nu,\bar\nu)}_i = -\frac{\gamma^{(\nu,\bar\nu)}}{2}\,,
\eeq
we arrive at \Eq{nudisprelimg}.

\section{Damping term in an $f\phi$ background}
\label{sec:dampingtermcalculation}

\subsection{Integral expressions for $\Sigma_i$ and $V^\mu_i$}

The diagram in \Fig{fig:sigma12fphi} gives the expression
\beq
\label{Sigma12diagram}
-i\Sigma_{12}(k) =
(i\lambda)(-i\lambda^\ast)\int\frac{d^4p}{(2\pi)^4}
i\Delta^{(\phi)}_{21}(q)RiS^{(f)}_{12}(p)L \,,
\eeq
where
\beq
q = p - k\,.
\eeq
The various components of the propagator matrices are
\beqa
S^{(f)}_{12}(p) & = & 2\pi i\delta(p^2 - m^2_f)
\left[\eta_F(p,\alpha_f) - \theta(-p\cdot u)\right](\lslash{p} + m_f)
\,,\nonumber\\
\Delta^{(\phi)}_{21}(q) & = & -2\pi i\delta(q^2 - m^2_\phi)
\left[\eta_B(q,\alpha_\phi) + \theta(q\cdot u)\right]\,,
\eeqa
where
\beqa
\eta_F(p,\alpha_f) & = & \theta(p\cdot u)n_F(x_f) +
\theta(-p\cdot u)n_F(-x_f)\,,\nonumber\\
\eta_B(q,\alpha_\phi) & = & \theta(q\cdot u)n_B(x_\phi) +
\theta(-q\cdot u)n_B(-x_\phi)\,,
\eeqa
with
\beqa
x_f & = & \beta p\cdot u - \alpha_f\,,\nonumber\\
x_\phi & = & \beta q\cdot u - \alpha_\phi \,
\eeqa
and $\theta$ is the step function. Further, $n_{F,B}$ are the fermion
and boson distribution functions already defined in \Eqs{nzfermion}{nzboson}.
%
%
It is useful to recall that the parameters $\alpha_{f,\phi}$ and
the corresponding one for the neutrinos, $\alpha$, must
satisfy
\beq
\alpha + \alpha_\phi = \alpha_f\,,
\eeq
by virtue of the $\bar f\nu\phi$ interaction. Therefore, together
with momentum conservation, the following useful relation holds
between $x_{f,\phi}$ and $x$,
\beq
\label{xrel}
x + x_\phi = x_f\,.
\eeq
Then, using identities of the $n_{F,B}$ functions as well
as \Eq{xrel}, after some manipulations we can write
\beqa
\label{DeltaSigma}
\Delta^{(\phi)}_{21}(q)S^{(f)}_{12}(p) & = &
(2\pi)^2(\lslash{p} + m_f)\delta(p^2 - m^2_f)
\delta(q^2 - m^2_\phi)\nonumber\\
&&\mbox{}\times n_F(x)\left(n_B(x_\phi) + n_F(x_f)\right)
\epsilon(q\cdot u)\epsilon(p\cdot u)\,,
\eeqa
and then from \Eq{Sigmaidef}
\beqa
\label{Sigmairewriten}
\Sigma_i & = &
-\frac{|\lambda|^2}{2}\int\frac{d^4p}{(2\pi)^3}\frac{d^4q}{(2\pi)^3}
(2\pi)^4 \delta^{(4)}(k + q - p)\delta(p^2 - m^2_f)\delta(q^2 - m^2_\phi)
\nonumber\\
&&\mbox{}\times
\left(n_B(x_\phi) + n_F(x_f)\right)
\epsilon(q\cdot u)\epsilon(p\cdot u)\lslash{p}L\,,
\eeqa
where $\epsilon(z) = \theta(z) - \theta(-z)$ with $\theta$ being the step
function. In writing \Eq{Sigmairewriten}, only the term of $S^{(f)}_{12}$
proportional to $\lslash{p}$ contributes due to chirality, 
and we have inserted the factor $d^4q\,\delta^{(4)}(k + q - p)$
for convenience in the manipulations that follow.

The integrations over $p^0$ and $q^0$ can be carried out with the help
of the $\delta$ functions, and with some change of variables,
\beqa
\label{Sigmairewriten2}
\Sigma_i & = & -\frac{|\lambda|^2}{2}
\int\frac{d^3p}{(2\pi)^3 2E}\frac{d^3q}{(2\pi)^3 2\Omega}
(2\pi)^4\left\{
\delta^{(4)}(k + q - p)(f_\phi + f_f)\right.\nonumber\\
&&\mbox{} +
\delta^{(4)}(k + p - q)(f_{\bar\phi} + f_{\bar f})\nonumber\\
&&\mbox{} +
\delta^{(4)}(k - q - p)(1 + f_{\bar\phi} - f_f)\nonumber\\
&&\mbox{} +
\left.\delta^{(4)}(k + q + p)(1 + f_{\phi} - f_{\bar f})
\right\}\lslash{p}L\,,
\eeqa
where
\beqa
E & = & \sqrt{\vec p^{\;2} + m^2_f}\,,\nonumber\\
\Omega & = & \sqrt{\vec q^{\;2} + m^2_\phi}\,,
\eeqa
and
\beqa
\label{standarddistributions}
f_{f,\bar f} & = & \frac{1}{e^{\beta E \mp \alpha_f} + 1}\,,\nonumber\\
f_{\phi,\bar\phi} & = & \frac{1}{e^{\beta \Omega \mp \alpha_\phi} - 1}\,.
\eeqa
\Eq{Sigmairewriten2} expresses the contributions to $\Sigma_i$ in terms
of the transition probabilities for the various processes in which
the neutrino may be annihilated or created,
\beqa
\label{reactions}
\nu + \phi & \leftrightarrow & f\nonumber\\
\nu + \bar f & \leftrightarrow & \bar\phi\nonumber\\
\nu & \leftrightarrow & f + \bar\phi\nonumber\\
\nu + \bar f + \phi & \leftrightarrow & 0\,.
\eeqa
This correspondence can be made more evident by rewriting the factors
that contain the distribution functions, for example,
\beq
f_\phi + f_f = f_\phi(1 - f_f) + f_f(1 + f_\phi)\,,
\eeq
and similarly with the other terms. The last two process
listed in \Eq{reactions} are forbidden by the kinematics,
assuming the vacuum dispersion relations for the particles and
$m_\nu = 0$. Therefore only the first two are viable.
However, as will be discussed in detail below,
the kinematic conditions under which they occur are different
and therefore not both are possible at the same time.
Thus, identifying $V^\mu_i$ by writing
\beq
\Sigma_i = R\lslash{V}_i L\,,
\eeq
then
\beqa
\label{Vmui}
V^\mu_i(\omega,\vec\kappa) & = & -\frac{|\lambda|^2}{2}
\int\frac{d^3p}{(2\pi)^3 2E}\frac{d^3q}{(2\pi)^3 2\Omega}
(2\pi)^4\left\{
\delta^{(4)}(k + q - p)(f_\phi + f_f)\right.\nonumber\\
&&\mbox{} +
\left.\delta^{(4)}(k + p - q)(f_{\bar\phi} + f_{\bar f})\right\} p^\mu\,.
\eeqa

We have considered above the case in which the neutrino propagates
in the presence of only one background, in this case composed of
$f$ and $\phi$ particles. In this case the distribution functions
that appear in \Eq{Vmui} have the standard isotropic form given
in \Eq{standarddistributions}. In a more general situation the medium
can consist of various superimposed backgrounds, and in principle
moving relative to each other with their own velocity four-vector
$u^\mu_s$. For example, we can envisage a medium composed of a normal
matter background, with a velocity four-vector $u^\mu$,
and which we can take to be at rest, plus a \emph{stream} background medium
of $f$ and $\phi$ particles with some velocity four-vector
$u^{\prime\,\mu} = (u^{\prime\,0},\vec u^{\,\prime})$.
In such cases, the distribution functions
that appear in \Eq{Vmui} would have the form
\beqa
\label{streamdistributions}
f^\prime_{f,\bar f} & = &
\frac{1}{e^{\beta p\cdot u^\prime \mp \alpha^\prime_f} + 1}\,,\nonumber\\
f^\prime_{\phi,\bar\phi} & = &
\frac{1}{e^{\beta q\cdot u^\prime \mp \alpha^\prime_\phi} - 1}\,,
\eeqa
which are not isotropic. In order to keep this possibility open,
we consider the evaluation of \Eq{Vmui} in the general case,
without assuming isotropy of the distribution functions,
and only consider the special isotropic case at the end.

\subsection{Formula for the damping term}

From \Eq{Vmui},
\beq
\label{Vmuicdotn}
V_i(\kappa,\vec\kappa)\cdot n = -\left(\frac{|\lambda|^2}{16\pi^2\kappa}\right)
\left[
\Delta^2_{f\phi}(K^\prime_\phi + K^\prime_f) +
\Delta^2_{\phi f}(L^\prime_{\bar\phi} + L^\prime_{\bar f})
\right]\,,
\eeq
where
\beqa
\label{KLprimedef}
K^\prime_\phi & = & \int\frac{d^3p}{2E}\frac{d^3q}{2\Omega}
\delta^{(4)}(k + q - p)f_\phi(\vec q)\,,\nonumber\\
K^\prime_f & = & \int\frac{d^3p}{2E}\frac{d^3q}{2\Omega}
\delta^{(4)}(k + q - p)f_f(\vec p)\,,\nonumber\\
L^\prime_{\bar\phi} & = & \int\frac{d^3p}{2E}\frac{d^3q}{2\Omega}
\delta^{(4)}(k + p - q)f_{\bar\phi}(\vec q)\,,\nonumber\\
L^\prime_{\bar f} & = & \int\frac{d^3p}{2E}\frac{d^3q}{2\Omega}
\delta^{(4)}(k + p - q)f_{\bar f}(\vec p)\,,
\eeqa
with
\beq
\label{Deltafphidef}
\Delta^2_{f\phi} = m^2_f - m^2_\phi\,,
\eeq
and
\beq
\label{Deltaphifdef}
\Delta^2_{\phi f} = m^2_\phi - m^2_f\,.
\eeq
In obtaining \Eq{Vmuicdotn} from \Eq{Vmui}
the factors $p\cdot n$ that would appear in the integrand
can be reduced as follows using the momentum conservation delta functions.
Consider the $K^\prime$ integrals, in which case
\beq
k + q = p\,.
\eeq
Then
\beq
(p - k)^2 = q^2 \Rightarrow 2p\cdot k = m^2_f - m^2_\phi\,,
\eeq
and using $k^\mu = \kappa n^\mu$,
\beq
p\cdot n = \frac{\Delta^2_{f\phi}}{2\kappa}\,.
\eeq
Similarly, in the case of the $L^\prime$ integrals,
the condition is $k + p = q$,
%
%
and the corresponding relation is
\beq
p\cdot n = \frac{\Delta^2_{\phi f}}{2\kappa}\,.
\eeq
This reduction is not possible in the case of the dispersive part
of the potential $V_r(\kappa,\vec\kappa)\cdot n$, therefore in
that case the factor $p\cdot n$ (which can be expressed in the form
$E(1 - \vec v\cdot\hat\kappa)$) remains in the integrand.
We consider the $K^\prime,L^\prime$ integrals one by one,
using the results for the generic integral $I^\prime$ given
in \Appendix{appendix:KLprime}. Although we are not indicating it explicitly,
it should be remembered that in these integrals we are
setting $\omega = \kappa$.

For $K^\prime_{\phi,f}$ the contributing process is
$\nu + \phi \leftrightarrow f$.
We then use the result given in \Eq{I12general-form12} with
the identification $1\rightarrow \phi, 2\rightarrow f$.
Thus, provided that
\beq
m_f > m_\phi\,,
\eeq
then
\beqa
\label{Kprimeresult}
K^\prime_\phi & = & \int\frac{d^3q}{2\Omega}\frac{1}{2\kappa|\vec q|}
\delta\left(\hat\kappa\cdot\hat q - \cos\theta^{(I)}_\phi\right)
\theta\left[\Omega - \Omega^{(I)}_{min}\right]f_\phi(\vec q)
\,,\nonumber\\
K^\prime_f & = & \int\frac{d^3p}{2E}\frac{1}{2\kappa|\vec p|}
\delta\left(\hat\kappa\cdot\hat p - \cos\theta^{(I)}_f\right)
\theta\left[E - E^{(I)}_{min}\right]f_f(\vec p)\,,
\eeqa
where
\beqa
\cos\theta^{(I)}_\phi & = &
\frac{2\kappa\Omega - \Delta^2_{f\phi}}{2\kappa|\vec q|}\,,\nonumber\\
\cos\theta^{(I)}_{f} & = &
\frac{2\kappa E - \Delta^2_{f\phi}}{2\kappa|\vec p|}\,,
\eeqa
and
\beqa
\label{Exmin}
%
\Omega^{(I)}_{min} & = & \frac{\Delta^2_{f\phi}}{4\kappa} +
\frac{\kappa m^2_\phi}{\Delta^2_{f\phi}}\,,\nonumber\\
E^{(I)}_{min} & = & \frac{\Delta^2_{f\phi}}{4\kappa} +
\frac{\kappa m^2_f}{\Delta^2_{f\phi}}\,,
\eeqa
with $\Delta^2_{f\phi}$ given in \Eq{Deltafphidef}.
Similarly, for $L^\prime_{\bar\phi,\bar f}$
the contributing process is $\nu + \bar f \leftrightarrow \bar\phi$,
and we use \Eq{I12general-form12} with the identifyication
$1\rightarrow \bar f, 2\rightarrow \bar\phi$. Thus, provided
\beq
m_\phi > m_f\,,
\eeq
we obtain in this case,
\beqa
\label{Lprimeresult}
L^\prime_{\bar\phi} & = &
\int\frac{d^3q}{2\Omega}\frac{1}{2\kappa|\vec q|}
\delta\left(\hat\kappa\cdot\hat q - \cos\theta^{(II)}_\phi\right)
\theta\left[\Omega - \Omega^{(II)}_{min}\right]
f_{\bar\phi}(\vec q)\,,\nonumber\\
L^\prime_{\bar f} & = & \int\frac{d^3p}{2E}\frac{1}{2\kappa|\vec p|}
\delta\left(\hat k\cdot\hat p - \cos\theta^{(II)}_f\right)
\theta\left[E - E^{(II)}_{min}\right]
f_{\bar f}(\vec p)\,,
\eeqa
where
\beqa
\cos\theta^{(II)}_\phi & = &
\frac{2\kappa\Omega - \Delta^2_{\phi f}}{2\kappa|\vec q|}\,,
\nonumber\\
\cos\theta^{(II)}_f & = &
\frac{2\kappa E - \Delta^2_{\phi f}}{2\kappa|\vec p|}\,,
\eeqa
and
\beqa
\label{Ebarxmin}
%
\Omega^{(II)}_{min} = \frac{\Delta^2_{\phi f}}{4\kappa} +
\frac{\kappa m^2_\phi}{\Delta^2_{\phi f}}\,,\nonumber\\
E^{(II)}_{min} = \frac{\Delta^2_{\phi f}}{4\kappa} +
\frac{\kappa m^2_f}{\Delta^2_{\phi f}}\,,
\eeqa
with $\Delta^2_{\phi f}$ in \Eq{Deltaphifdef}.

Using these results, from \Eq{nudisprelimg-simple}
the damping rate for the neutrino is then
%
%
\beq
\label{gammanuanisotropic}
\frac{\gamma^{(\nu)}(\vec\kappa)}{2} =
\frac{|\lambda|^2}{16\pi^2\kappa}
\left\{\begin{array}{ll}
\Delta^2_{f\phi}(K^\prime_\phi + K^\prime_f) &
(m_f > m_\phi),\; (\nu + \phi \leftrightarrow f)\\[12pt]
\Delta^2_{\phi f}(L^\prime_{\bar\phi} + L^\prime_{\bar f}) &
(m_\phi > m_f),\; (\nu + \bar f\leftrightarrow \bar\phi)
\end{array}\right.
\eeq
with $K^\prime_{f,\phi},L^\prime_{\bar f,\bar\phi}$ given
in \Eqs{Kprimeresult}{Lprimeresult}. We have indicated in parenthesis
the kinematic condition on the masses and, for reference purposes,
the contributing process. For the antineutrino the relevant quantity is
$V_i(-\kappa,-\vec\kappa)\cdot n$. From \Eq{Vmui}, in this case we obtain
\beq
V_i(-\kappa,-\vec\kappa)\cdot n =
-\left(\frac{|\lambda|^2}{16\pi^2\kappa}\right)\left[
\Delta^2_{f\phi}(\bar K^\prime_\phi + \bar K^\prime_f) +
\Delta^2_{\phi f}(\bar L^\prime_{\bar\phi} + \bar L^\prime_{\bar f})
\right]\,,
\eeq
where $\bar K^\prime_{f,\phi},\bar L^\prime_{\bar f,\bar\phi}$ are defined
as $K^\prime_{f,\phi},L^\prime_{\bar f,\bar\phi}$ in \Eq{KLprimedef},
but making the substitutions
\beq
\label{KLbarsubstitutions}
f_f \leftrightarrow f_{\bar f}\,,\qquad f_\phi \leftrightarrow f_{\bar\phi}\,.
\eeq
Correspondingly, the final formulas for
$\bar K^\prime_{f,\phi},\bar L^\prime_{\bar f,\bar\phi}$
can be obtained from \Eqs{Kprimeresult}{Lprimeresult} by
making the same substitutions. Thus, from \Eq{nudisprelimg-simple},
for the antineutrino
\beq
\label{gammabarnuanisotropic}
\frac{\gamma^{(\bar\nu)}(\vec\kappa)}{2} =
\frac{|\lambda|^2}{16\pi^2\kappa}
\left\{\begin{array}{ll}
\Delta^2_{f\phi}(\bar K^\prime_\phi + \bar K^\prime_f) &
(m_f > m_\phi),\; (\bar\nu + \bar\phi\leftrightarrow \bar f)\\[12pt]
\Delta^2_{\phi f}(\bar L^\prime_{\bar\phi} + \bar L^\prime_{\bar f}) &
(m_\phi > m_f),\; (\bar\nu + f \leftrightarrow \phi)\end{array}\right.
\eeq

\subsection{Evaluation of the damping term for an isotropic $f\phi$ background}

When the momentum distributions have the standard form the $K^\prime,L^\prime$
integrals can be readily evaluated. The results can be read off
the generic result given in \Eq{Iaisotropic} with the same identifications
as made above, namely $1\rightarrow \phi, 2\rightarrow f$ and
$1\rightarrow \bar f, 2\rightarrow \bar\phi$
for the $K^\prime$ and $L^\prime$ integrals, respectively.
Thus, using \Eq{Iaisotropic} in the manner we have indicated,
we obtain from \Eq{gammanuanisotropic} the following result
for the damping term, depending on whether $m_f > m_\phi$ or $m_\phi > m_f$,
\beq
\label{gammanuisotropicresult}
\frac{\gamma^{(\nu)}(\kappa)}{2} = \frac{|\lambda|^2}{32\pi\kappa^2\beta}
\left\{
\begin{array}{ll}
\Delta^2_{f\phi}\left[
\log\left(1 + e^{-\beta E^{(I)}_{min} + \alpha_f}\right) -
\log\left(1 - e^{-\beta \Omega^{(I)}_{min} + \alpha_\phi}\right)
\right] & (m_f > m_\phi) \\[12pt]
\Delta^2_{\phi f}\left[
\log\left(1 + e^{-\beta E^{(II)}_{min} - \alpha_f}\right) -
\log\left(1 - e^{-\beta \Omega^{(II)}_{min} - \alpha_\phi}\right)
\right] & (m_\phi > m_f)
\end{array}\right.
\eeq
The corresponding formulas for
$\frac{\gamma^{(\bar\nu)}(\kappa)}{2}$ are
obtained by making the substitutions
\beq
\alpha_{f,\phi} \rightarrow - \alpha_{f,\phi}\,,
\eeq
in \Eq{gammanuisotropicresult}. In particular,
if the background is particle-antiparticle symmetric,
so that the chemical potentials $\alpha_{f,\phi}$ are zero, then
$\gamma^{(\bar\nu)}(\kappa)$ and $\gamma^{(\nu)}(\kappa)$ are equal,
as expected on general grounds based on the $CPT$ symmetry of the
background in that case.

\section{Discussion}
\label{sec:discussion}

We illustrate some particular features of \Eq{gammanuisotropicresult}
by considering specific situations.

\subsection{Degenerate $f$ background}

For example,
let us consider the case in which the scalar particle is so heavy
that it is not present in the background. To be
specific let us take $m_\phi \gg m_f$. In this example the contributing
process is $\nu + \bar f \leftrightarrow \bar\phi$.
Then from \Eq{gammanuisotropicresult},
\beq
\frac{\gamma^{(\nu)}(\kappa)}{2} =
\frac{|\lambda|^2 m^2_\phi}{32\pi\kappa^2\beta}
\log\left(1 + e^{-\beta \epsilon^{(II)}_{min} - \alpha_f}\right)\,.
\eeq
where, from \Eq{Ebarxmin}, we have approximated
\beq
\label{Ebarxmin-ex1}
E^{(II)}_{min} \simeq \epsilon^{(II)}_{min} \equiv
\frac{m^2_\phi}{4\kappa} + \frac{\kappa m^2_f}{m^2_\phi}
\eeq
in the present case.
Let us further assume that the $\bar f$ fermions form a completely
degenerate gas and there are no $f$ fermions. Then setting
\beq
\alpha_f = -\beta E_F\,,
\eeq
where $E_F$ is the Fermi energy, and taking the degenerate limit
($\beta \rightarrow \infty$),
\beq
\label{gammanuex1}
\frac{\gamma^{(\nu)}(\kappa)}{2} = \frac{|\lambda|^2 m^2_\phi}{32\pi\kappa^2}
\left(E_F - \epsilon^{(II)}_{min}\right)
\theta\left(E_F - \epsilon^{(II)}_{min}\right)\,,
\eeq
while $\gamma^{(\bar\nu)}(\kappa) = 0$. The step function in \Eq{gammanuex1}
implies that $\gamma^{(\nu)}(\kappa)$ is non-zero if $\kappa$ lies
in the range such that
\beq
\label{kappalimitex1}
\frac{1}{2}(E_F - p_F) < \frac{\kappa m^2_f}{m^2_\phi} <
\frac{1}{2}(E_F + p_F)\,,
\eeq
or it is zero otherwise\footnote{%
To prove \Eq{kappalimitex1} we rewrite the condition
$E_F > \epsilon^{(II)}_{min}$ in the form
\beq
\label{conditionforx}
\frac{1}{4x} + x < \frac{E_F}{m_f}\,,
\eeq
where
\beq
x \equiv \frac{\kappa m_f}{m^2_\phi}\,.
\eeq
Now let us solve the equation
\beq
\label{eqforx}
\frac{1}{4x} + x = a\,,
\eeq
where $a$ is, at the moment, unspecified. The solutions are
\beq
x_{\pm}(a) = \frac{a}{2} \pm \frac{1}{2}\sqrt{a^2 - 1}\,.
\eeq
The functions $x_{\pm}(a)$ are equal at $a = 1$, and as
$a$ increases $x_{+}$ increases while $x_{-}$ decreases.
Since these functions, by construction, satisfy \Eq{eqforx},
then their value at any $a$ (subject to $a > 1$) satisfies
\Eq{conditionforx} for all $a < E_F/m_f$. It then follows that
all the values of $x$ that lie between $x_{-}(E_F/m_f)$ and
$x_{+}(E_F/m_f)$ satisfy \Eq{conditionforx}, while the values outside
that range correspond to $a > E_F/m_f$ and therefore will violate it.
Using the fact that
\beq
x_{\pm}(E_F/m_f) = \frac{1}{2}\left(\frac{E_F}{m_f}\right)
\pm \frac{1}{2}\sqrt{\left(\frac{E_F}{m_f}\right)^2 - 1} =
\frac{1}{2}\left(\frac{E_F}{m_f}\right)
\pm \frac{1}{2}\frac{p_F}{m_f}\,,
\eeq
proves \Eq{kappalimitex1}.}
On the other hand, if we assume that it is the $f$ fermions that
form a completely degenerate gas and there are no $\bar f$ fermions,
then we have $\gamma^{(\nu)}(\kappa) = 0$ while
$\gamma^{(\bar\nu)}(\kappa)$ is given by a formula
identical to \Eq{gammanuex1}, but in this case $E_F$ referring to the
Fermi energy of the $f$ fermion gas and $\alpha_f = \beta E_F$.

\subsection{Classical relativistic $f$ and $\bar f$ gas}

In contrast, if we take a neutral ($\alpha_f = 0$) gas of fermions,
and assume that it can be considered in the classical regime
then we obtain
\beqa
\label{gammanuex2}
\frac{\gamma^{(\nu)}(\kappa)}{2} = \frac{\gamma^{(\bar\nu)}(\kappa)}{2} \simeq
\frac{|\lambda|^2 m^2_\phi}{32\pi\kappa^2\beta}e^{-\beta\epsilon^{(II)}_{min}}\,,
\eeqa
where $\epsilon^{(II)}_{min}$ has been defined in \Eq{Ebarxmin-ex1}.
In this case, the damping has a maximum value at the value of $\kappa$
determined by the condition
\beq
\label{stationarycond-ex2}
\frac{d\epsilon^{(II)}_{min}}{d\kappa} = -\frac{2}{\beta\kappa}\,,
\eeq
where
\beq
\frac{d\epsilon^{(II)}_{min}}{d\kappa} = -\frac{m^2_\phi}{4\kappa^2} +
\frac{m^2_f}{m^2_\phi}\,.
\eeq
As an example, suppose that the conditions, to be determined below,
are such that the term $m^2_f/m^2_\phi$ in \Eq{stationarycond-ex2}
can be neglected. This gives
\beq
\label{stationarycond-ex2-kappa}
\kappa = \frac{\beta m^2_\phi}{8}\,.
\eeq
Notice that for this value of $\kappa$,
\beq
\frac{2}{\beta\kappa} = \frac{m^2_\phi}{4\kappa^2} =
\frac{16}{\beta^2 m^2_\phi}\,,
\eeq
so that indeed the $m^2_f/m^2_\phi$ term can be neglected 
if the $f,\bar f$ gas is ultra-relativistic ($\beta m_f \ll 1$).
Consequently the maximum value of the damping is
\beq
\label{maxdampingex2}
\frac{\gamma^{(\nu)}}{2} = \frac{\gamma^{(\bar\nu)}}{2} = 
\frac{2|\lambda|^2}{\pi m^2_\phi \beta^3}e^{-2}\,,
\eeq
which occurs at the value of $\kappa$ given in \Eq{stationarycond-ex2-kappa}.
For other values of $\kappa$ the damping drops exponentially.

\subsection{Classical relativistic $\phi$ background}
  
We now consider the opposite case, namely $m_f > m_\phi$ and assume that the
conditions are such that there are no fermions in the background.
In addition we assume that the scalar $\phi$ is its own antiparticle
and therefore $\alpha_\phi = 0$. In this case the contributing
process is $\nu + \phi \leftrightarrow f$. From \Eq{gammanuisotropicresult},
\beqa
\frac{\gamma^{(\nu)}(\kappa)}{2} = \frac{\gamma^{(\bar\nu)}(\kappa)}{2} & = &
-\frac{|\lambda|^2 m^2_f}{32\pi\kappa^2\beta}
\log\left(1 - e^{-\beta \omega^{(I)}_{min}}\right)\nonumber\\
& = & \frac{|\lambda|^2 m^2_f}{32\pi\kappa^2\beta}
e^{-\beta \omega^{(I)}_{min}}\,,
\eeqa
where in the last line we have approximated
\beq
\Omega^{(I)}_{min} \simeq \omega^{(I)}_{min} \equiv
\frac{m^2_f}{4\kappa} + \frac{\kappa m^2_\phi}{m^2_f}\,,
\eeq
and assumed that the $\phi$ background can be treated within the classical
approximation. Furthermore, if we consider a relativistic $\phi$ gas, then
by carrying out steps similar to those leading to
\Eq{maxdampingex2}, we obtain in this case the maximum value
of the damping
\beq
\label{maxdampingex3}
\frac{\gamma^{(\nu)}}{2} = \frac{\gamma^{(\bar\nu)}}{2} = 
\frac{2|\lambda|^2}{\pi m^2_f \beta^3}e^{-2}\,,
\eeq
at the momentum
\beq
\label{stationarycond-ex3-kappa}
\kappa = \frac{\beta m^2_f}{8}\,.
\eeq

\section{Damping term in a $\nu\phi$ background}

The results we have obtained in the previous sections can
be adapted to the case of a neutrino propagating in a neutrino
and scalar background. For example, we can assume that a sterile ($\nu_{s}$)
and an active neutrino have an interaction of the form
\beq
\label{Lnusnuphi}
L_{int}= \lambda^{(\nu\nu_s\phi)}{\bar \nu}^c_{Rs} {\nu}_L\phi + h.c.
\eeq
and consider the cases of either $\nu$ or $\nu_s$ propagating
in a background of the other plus $\phi$. With the appropriate identifications
the formulas we have presented can be used in either case,
with the provision that the propagating neutrino is considered to be massless.

\begin{figure}
\begin{center}
\epsfig{file=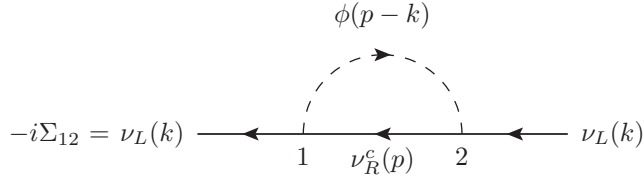,bbllx=73,bblly=329,bburx=534,bbury=443}
\end{center}
\caption{
  One-loop diagram for the $\Sigma_{12}$ element of the neutrino thermal
  self-energy in a $\nu\phi$ background.
  \label{fig:sigma12nuphi}
}
\end{figure}

In the rest of this section we consider another case, namely the
damping effects in a neutrino and scalar background, due to the diagonal
interaction given in \Eq{Lnunuphi}. In the relevant diagram for this case,
shown in \Fig{fig:sigma12nuphi}, the internal fermion that corresponds to the
$f_R$ fermion line in \Fig{fig:sigma12fphi} is the antineutrino $\nu^c_R$.
Thus, the formulas obtained in
\Sections{sec:dampingtermcalculation}{sec:discussion}
can be adapted to this case by identifying
$f,\bar f\rightarrow \bar\nu,\nu$ in the labels of the various
physical quantities that refer to the background particles,
such as the chemical potentials. We restrict ourselves here
to the case of isotropic $\nu$ and $\phi$ backgrounds,
in which case we can use the formulas given in \Eq{gammanuisotropicresult}
as the starting point. Since the background neutrino
is the same as the initial neutrino, which have taken to be massless, we
use the formula for $m_\phi > m_f$ (which corresponds to
the process $\nu + \nu \leftrightarrow \bar\phi$). Therefore
\beq
\label{gammanunuphiresult}
\frac{\gamma^{(\nu)}(\kappa)}{2} =
\frac{|\lambda^{(\nu\nu\phi)}|^2}{32\pi\kappa^2\beta}
\Delta^2_{\phi\nu}\left[
\log\left(1 + e^{-\beta E^{(II)}_{min} + \alpha_\nu}\right) -
\log\left(1 - e^{-\beta \Omega^{(II)}_{min} - \alpha_\phi}\right)
\right]\,,
\eeq
where we have set $\alpha_f \rightarrow \alpha_{\bar\nu} = -\alpha_\nu$,
as we have indicated above. Furthermore, borrowing from
\Eq{Ebarxmin} and remembering that in this case $m_\nu \rightarrow 0$,
\beqa
\Omega^{(II)}_{min} & = & \kappa + \frac{m^2_\phi}{4\kappa}\,,\nonumber\\
E^{(II)}_{min} & = & \frac{m^2_\phi}{4\kappa}\,.
\eeqa
The corresponding formula for $\gamma^{(\bar\nu)}(\kappa)$ is given by
making the substitutions
\beq
\alpha_\nu \rightarrow -\alpha_\nu,\;
\alpha_\phi \rightarrow -\alpha_\phi\,,
\eeq
in \Eq{gammanunuphiresult}.

The formula in \Eq{gammanunuphiresult} and the corresponding one for
$\gamma^{(\bar\nu)}(\kappa)$ can be used to study different cases,
for example, different conditions of the background gases,
that can be useful in specific situations.
For illustrative purposes let us consider the particular case in which
the background density of the $\phi$ particles can be neglected.
Assuming further that the neutrino gas can be treated in the classical regime,
then we can use \Eq{maxdampingex2} and obtain
\beq
\frac{\gamma^{(\nu)}}{2} = \frac{\gamma^{(\bar\nu)}}{2} = 
\frac{2|\lambda^{(\nu\nu\phi)}|^2 T^3_\nu}{\pi m^2_\phi}e^{-2}\,,
\eeq
where we are denoting by $T_\nu$ the temperature of the
neutrino gas. Here we are assuming that the neutrino chemical potential
is sufficiently small that we can use the damping formula with
$\alpha_\nu \sim 0$. In order to assess the possible relative importance
of the damping effects in this case, we recall from \Ref{Nieves:2018vxl}
the following result for the real part of the dispersion relation
for this same example case,
\beq
\omega^{(\nu,\bar\nu)}_r(\vec\kappa) =
\kappa + V^{(\nu,\bar\nu)}_{eff}(\vec\kappa)\,,
\eeq
where
\beq
\label{Veffnuphi+Z}
V^{(\nu,\bar\nu)}_{eff}(\vec\kappa) = \pm 
\left(2\sqrt{2}G_F - \frac{|\lambda^{(\nu\nu\phi)}|^2}{2m^2_\phi}\right)
(n_\nu - n_{\bar\nu}) + O(1/m^4_\phi)\,.
\eeq
Therefore, the importance of the damping term depends on the magnitude of
\beq
\frac{|\lambda^{(\nu\nu\phi)}|^2 T^3_\nu}{m^2_\phi}\,,
\eeq
relative to
\beq
\frac{g^2}{m^2_Z}(n_\nu - n_{\bar\nu})\,,
\eeq
and 
\beq
\frac{|\lambda^{(\nu\nu\phi)}|^2}{m^2_\phi}(n_\nu - n_{\bar\nu})\,.
\eeq

As pointed out in \Ref{Nieves:2018vxl},
requiring that the $\phi$ exchange contribution to
$\omega^{(\nu,\bar\nu)}_r(\vec\kappa)$ be smaller than the
standard $Z$-exchange contribution implies that
\beq
m_\phi > \frac{|\lambda^{(\nu\nu\phi)}|}{g}m_Z\,.
\eeq
For example, if $|\lambda^{(\nu\nu\phi)}| \sim O(1)$, this would imply that
$m_\phi$ must be larger than $\sim 3m_Z$. On the other hand, in order to have
$m_\phi \sim O(GeV)$, a small coupling
$|\lambda^{(\nu\nu\phi)}| \sim 10^{-2} - 10^{-3}$ would be required.

However, as this example shows, the damping term can be as large
as the $\nu\phi$ contribution to $V^{(\nu,\bar\nu)}_{eff}$.
In fact, the relative importance of the damping term
may be more significant if the neutrino-antineutrino asymmetry
in the background is small, because the $O(1/m^2_Z)$ and $O(1/m^2_\phi)$
contributions to $V^{(\nu,\bar\nu)}_{eff}$ are suppressed in that case, while
the damping term is not.

\section{Conclusions}
\label{sec:conclusions}

The neutrino interactions with scalar particles via
couplings of the form $\lambda^{(\nu\nu\phi)}\bar\nu^c_{R}\nu_{L}\phi$
can produce nonstandard contributions to the neutrino index
of refraction and effective potential when the neutrino propagates
in a neutrino background. This occurs in the environment of a supernova,
and in the Early Universe hot plasma before neutrino decoupling.
Motivated by this, we had previously\cite{Nieves:2018vxl}
considered a simple model in which the neutrino interacts with a scalar
and a fermion background with a coupling of the form
$\lambda\bar f_R\nu_L\phi$. In \Ref{Nieves:2018vxl} we focused exclusively
on the calculation of the real part of the dispersion relation and
effective potential for a neutrino propagating in a thermal background of
those particles.

In the present work we have complemented those calculations
by considering the damping terms in the neutrino effective potential and
index of refraction. The damping terms arise as a consequence ofo processes
such as $\nu + \phi \leftrightarrow f$ and
$\nu + \bar f\leftrightarrow \bar\phi$
that occur when the neutrino propagates in the $f\phi$ background,
depending on the kinematics conditions. Specifically,
we calculated the imaginary part of the neutrino self-energy, from which
the damping terms in the neutrino effective potential
and dispersion relations were determined.
The formulas for the case of the neutrino and scalar background,
were then obtained as the special case in which $f_R \rightarrow \nu^c_R$.

The results are useful in the context of Dark Matter-neutrino
interaction models in which the scalar and/or fermion constitute the
dark-matter. They are also applicable to the situations
mentioned above in which the fermion background is a neutrino background,
including, for example, a sterile neutrino
interacting with a neutrino background or the other way around.
We obtained the expressions for the damping terms in the
neutrino effective potential that can be applied to different
situations and background conditions. For definiteness,
we obtained explicit formulas
for the damping terms in the case that the background particle momentum
distributions have the standard isotropic form, but for generality
and in order to present the results in a way that can be useful
to the situations described and others not considered here, the
generalization to the case of an anisotropic background was mentioned.

As specific application we considered the case in which
the scalar is sufficiently heavy such that their distribution
function can be neglected, and for definiteness
the neutrino gas can be treated in the classical regime.
As that particular example showed, the damping term can be as large
as the contribution to effective potential.
Moreover, the importance of the damping term
becomes more significant when the neutrino-antineutrino asymmetry
in the background is small, because the $O(1/m^2_Z)$ and $O(1/m^2_\phi)$
contributions to the effective potential are proportional to the
neutrino-antineutrino asymmetry and therefore are suppressed in that case,
while the damping term is not.

Despite the simplification and idealization we have made by
considering only the diagonal scalar-neutrino coupling,
the formulas and results we have presented
can serve as a guide when considering more realistic or
complicated situations, such as those involving more than one scalar
particle or off-diagonal neutrino couplings, and to determine whether the
damping effects are significant and must be considered or if they are
small and can be neglected. For example, with off-diagonal neutrino couplings
one neutrino can decay into another neutrino and a $\phi$ scalar
(the third process type listed
in \Eq{reactions}. The corresponding contribution
to the damping can be calculated in a manner similar to the calculations
we have presented.
  
The work of S. S. is partially supported by DGAPA-UNAM
(Mexico) Project No. IN103019.

\appendix

\section{Evaluation of integrals}
\label{appendix:KLprime}

Here we derive the results quoted in \Eqs{Kprimeresult}{Lprimeresult}.
Since the integrals involved have the generic form
\beq
\label{Iprime}
I^\prime(\omega,\vec\kappa) \equiv \int\frac{d^3p_1}{2E_1}\frac{d^3p_2}{2E_2}
\delta^{(4)}(k + p_1 - p_2)F(\vec p_1,\vec p_2)\,,
\eeq
we focus our attention for the moment on this kind of integral. We set
\beq
\omega  = \kappa\,,
\eeq
from the beginning since this is the appropriate limit for our purposes.
Rewriting it as
\beq
I^\prime(\kappa,\vec\kappa) = \int\frac{d^3p_1}{2E_1}d^4p_2
\delta(p^2_2 - m^2_2)\theta(p^0_2)
\delta^{(4)}(k + p_1 - p_2)F(\vec p_1,\vec p_2)\,,
\eeq
and doing the integral over $p_2$ first, we set
\beq
\label{p2}
p_2 = p_1 + k\,,
\eeq
and thus
\beqa
I^\prime(\kappa,\vec\kappa) & = &
\int\frac{d^3p_1}{2E_1}\delta[(k + p_1)^2 - m^2_2]\theta(\kappa + E_1)
F(\vec p_1,\vec p_1 + \vec\kappa)\nonumber\\
& = &  \int\frac{d^3p_1}{2E_1}\left(\frac{1}{2\kappa|\vec p_1|}\right)
F(\vec p_1,\vec p_1 + \vec\kappa)\delta(\hat\kappa\cdot\hat p_1 - \cos\theta_1)
\theta(\kappa + E_1)\,,
\eeqa
where
\beq
\label{costheta1}
\cos\theta_1 \equiv
\frac{2\kappa E_1 - \Delta^2_{21}}{2\kappa p_1}\,,
\eeq
with
\beq
\label{Delta21}
\Delta^2_{21} = m^2_2 - m^2_1\,.
\eeq
$\cos\theta_1$ must satisfy
\beq
\label{costheta1cond}
-1 \le \cos\theta_1 \le 1\,,
\eeq
which requires
\beq
\Delta^2_{21} > 0\,,
\eeq
and also determines the $E_1$ limits of integration that we denote by
$(E_1)_{min}$ and $(E_1)_{max}$. From the fact that
$\vec p_2 = \vec\kappa + \vec p_1$ together with \Eq{costheta1}
it follows that
\beq
\label{costheta2}
\hat\kappa\cdot\hat p_2 = \cos\theta_2 \equiv
\frac{2\kappa E_2 - \Delta^2_{21}}{2\kappa p_2}\,,
\eeq
where
\beq
\label{E2}
E_2 = E_1 + \kappa\,.
\eeq
In particular,
\beq
\label{E2minmaxdef}
(E_2)_{min,max} = (E_1)_{min,max} + \kappa\,.
\eeq

To find the limits implied by \Eq{costheta1} we solve that relation
for $E_1,p_1$ as functions of $\cos\theta_1$. We rewrite it as
\beq
\label{p1xeq}
E_1 = xp_1 + \xi m_1\,,
\eeq
where
\beqa
x & = & \cos\theta_1\,,\nonumber\\
\xi & = & \frac{\Delta^2_{21}}{2\kappa m_1}\,.
\eeqa
Then, putting $E_1 = \sqrt{p^2_1 + m^2_1}$ and solving \Eq{p1xeq} for
$p_1$ yields
\beq
p_1 = \frac{m_1}{y^2}\left\{x\xi \pm \sqrt{\xi^2 - y^2}\right\}\,,
\eeq
where
\beq
y = \sqrt{1 - x^2} = \sin\theta_1\,,
\eeq
and then
\beq
E_1 = \frac{m_1}{y^2}\left\{\xi \pm x\sqrt{\xi^2 - y^2}\right\}\,.
\eeq
These solutions for $E_1,p_1$ of course satisfy $E^2_1 = p^2_1 + m^2_1$,
but we must be sure that they satisfy
\beq
E_1, p_1 > 0\,,
\eeq
as well. There are two possibilities.
\subsubsection*{Case I: $\xi > 1 \qquad (\kappa < \frac{\Delta^2_{21}}{2m_1})$}

In this case we have
\beq
\sqrt{\xi^2 - y^2} > x\xi\,,
\eeq
and therefore the solution with the negative sign is not valid. Therefore
in this case the only solution is
\beqa
\label{p1E1solution1}
p_1 & = & \frac{m_1}{y^2}\left\{x\xi + \sqrt{\xi^2 - y^2}\right\}\,,\nonumber\\
E_1 & = & \frac{m_1}{y^2}\left\{\xi + x\sqrt{\xi^2 - y^2}\right\}\,.
\eeqa

\subsubsection*{Case II: $\xi < 1 \qquad (\kappa > \frac{\Delta^2_{21}}{2m_1})$}

In this case the factor $\sqrt{\xi^2 - y^2}$ implies that there is
a maximum value of $y$,
\beq
y_{max} = \xi\,,
\eeq
so that
\beq
\sqrt{\xi^2 - y^2} < x\xi\,,
\eeq
which in turn implies that the solution with the minus sign is also allowed.
Thus,
\beqa
\label{p1E1solution2}
p^{(\pm)}_1 & = & \frac{m_1}{y^2}\left\{x\xi \pm \sqrt{\xi^2 - y^2}\right\}\,,
\nonumber\\
E^{(\pm)}_1 & = & \frac{m_1}{y^2}\left\{\xi \pm x\sqrt{\xi^2 - y^2}\right\}\,,
\eeqa
where $p_1 > 0 \Rightarrow x > 0$.

In either case, for $\cos\theta \rightarrow 1$
(which implies $y \rightarrow 0$) we have
\beq
(E_1)_{max} = \left.E_1\right|_{x \rightarrow 1} = \infty\,.
\eeq
On the other hand, $(E_1)_{min}$ is given by
$\left.E_1\right|_{x \rightarrow -1}$ in Case I,
or $\left.E^{(-)}_1\right|_{x \rightarrow 1}$ in Case II. In either case,
\beqa
\label{summaryE1min}
(E_1)_{min} & = & \left\{
\frac{m_1}{y^2}\left[\xi - \left(1 - \frac{y^2}{2}\right)
\left(\xi - \frac{y^2}{2\xi}\right)\right]\right\}_{y \rightarrow 0}\nonumber\\
& = & m_1\left[\frac{\xi}{2} + \frac{1}{2\xi}\right]\nonumber\\
& = & \frac{\Delta^2_{21}}{4\kappa} + \frac{\kappa m^2_1}{\Delta^2_{21}}\,.
\eeqa
From \Eq{E2minmaxdef},
\beq
\label{summaryE2min}
(E_2)_{min} = \frac{\Delta^2_{21}}{4\kappa} +
\frac{\kappa m^2_2}{\Delta^2_{21}}\,.
\eeq

We can summarize the results as follows. The kinematic conditions require that
\beq
m_2 > m_1\,,
\eeq
and the integration over $p_1$ is restricted by the condition
\beq
E_1 \ge (E_1)_{min}\,,
\eeq
where $(E_1)_{min}$ is given in \Eq{summaryE1min}. Then
$I^\prime(\kappa,\vec\kappa)$ can be expressed in the form
\beq
\label{I12general-form1}
I^\prime(\kappa,\vec\kappa) =
\int\frac{d^3p_1}{2E_1}\left(\frac{1}{2\kappa|\vec p_1|}\right)
\delta(\hat\kappa\cdot\hat p_1 - \cos\theta_1)
\theta\left[E_1 - (E_1)_{min}\right]
F(\vec p_1,\vec p_1 + \vec\kappa)\qquad \mbox{(Form-1)}\,,
\eeq
to which we refer as the \emph{Form-1}.  It can be expressed in the
following equivalent form by carrying the integral over $p_1$ first,
\beq
\label{I12general-form2}
I^\prime(\kappa,\vec\kappa) =
\int\frac{d^3p_2}{2E_2}\left(\frac{1}{2\kappa|\vec p_2|}\right)
\delta(\hat\kappa\cdot\hat p_2 - \cos\theta_2)
\theta\left[E_2 - (E_2)_{min}\right]
F(\vec p_2 - \vec\kappa,\vec p_2)\qquad \mbox{(Form-2)}\,,
\eeq
where $\cos\theta_2$ and $(E_2)_{min}$ are given in
\Eqs{costheta2}{summaryE2min}, respectively.
We refer to \Eq{I12general-form2} as the \emph{Form-2}.
It then follows that if we define the integrals
\beq
\label{Iprime12}
I^\prime_{a}(\omega,\vec\kappa) \equiv
\int\frac{d^3p_1}{2E_1}\frac{d^3p_2}{2E_2}
\delta^{(4)}(k + p_1 - p_2)f_a(p_a) \qquad (a = 1,2)\,,
\eeq
then, using \Eqs{I12general-form1}{I12general-form2} for $a = 1,2$,
respectively, we get
\beq
\label{I12general-form12}
I^\prime_a(\kappa,\vec\kappa) =
\int\frac{d^3p_a}{2E_a}\left(\frac{1}{2\kappa|\vec p_a|}\right)
\delta(\hat\kappa\cdot\hat p_a - \cos\theta_a)
\theta\left[E_a - (E_a)_{min}\right] f_a(\vec p_a)\,.
\eeq
In the case that the momentum distributions have
the standard isotropic form [e.g., \Eq{standarddistributions}],
the expression for $I^\prime_a(\kappa,\vec\kappa)$
in \Eq{I12general-form12} reduces to
\beq
\label{Iaisotropic}
I^\prime_a(\kappa,\vec\kappa) = \frac{\pi}{2\kappa}I_a\,,
\eeq
where $I_a$ is the elementary integral
\beq
\label{Iaelementary}
I_a = \int^\infty_{(E_a)_{min}}dE_a\, f_a(E_a) = 
\left\{
\begin{array}{ll}
\frac{1}{\beta}\log
\left[1 + e^{-\beta (E_a)_{min} + \alpha_a}\right] & (F)\\[12pt]
-\frac{1}{\beta}\log
\left[1 - e^{-\beta (E_a)_{min} + \alpha_a}\right] & (B)
\end{array}\right.
\eeq
where the labels $F$ and $B$ stand for either the
fermion or boson distribution function, respectively.
\Eq{I12general-form12}, and its counterpart \Eq{Iaisotropic} in the isotropic
case, are the formulas we have used in the text.

For easy reference, we summarize here the following formulas
that enter in \Eqss{I12general-form1}{I12general-form2}{I12general-form12},
\beqa
\label{I12parametersummary}
(E_1)_{min} & = & \frac{\Delta^2_{21}}{4\kappa} +
\frac{\kappa m^2_1}{\Delta^2_{21}}\,,
\nonumber\\
\cos\theta_1 & = & \frac{2\kappa E_1 - \Delta^2_{21}}{2\kappa|\vec p_1|}\,,
\nonumber\\
(E_2)_{min} & = & \frac{\Delta^2_{21}}{4\kappa} +
\frac{\kappa m^2_2}{\Delta^2_{21}}\,,
\nonumber\\
\cos\theta_2 & = & \frac{2\kappa E_2 - \Delta^2_{21}}{2\kappa|\vec p_2|}\,,
\nonumber\\
\Delta^2_{21} & \equiv & m^2_2 - m^2_1\,.
\eeqa
%


\end{document}